# Resonantly driven singlet-triplet spin qubit in silicon

K. Takeda,[1] A. Noiri,[1] J. Yoneda,[1,*] T. Nakajima,[1] and S. Tarucha[1]

1. Center for Emergent Matter Science (CEMS), RIKEN, Wako-shi, Saitama, 351-0198, Japan

* Present address: School of Electrical Engineering and Telecommunications, University of New South Wales, Sydney, New South Wales 2052, Australia

Abstract

We report implementation of a resonantly driven singlet-triplet spin qubit in silicon. The qubit is defined by the two-electron anti-parallel spin states and universal quantum control is provided through a resonant drive of the exchange interaction at the qubit frequency. The qubit exhibits long $T_2^*$ exceeding 1 μs that is limited by dephasing due to the $^{29}$Si nuclei rather than charge noise thanks to the symmetric operation and a large micro-magnet Zeeman field gradient. The randomized benchmarking shows 99.6 % single gate fidelity which is the highest reported for singlet-triplet qubits.

Main text

Electron spins confined in semiconductor quantum dots (QDs) are attractive candidates for implementing scalable solid-state quantum computing [1]. Recent technical advances have enabled high-fidelity single- and two-qubit control for spin-1/2 qubits in this system [2–7]. While the spin-1/2 qubit is the most straightforward implementation of a spin qubit, there are a number of attempts to encode a qubit using more than one electron spins in multiple QDs to benefit from the increased degrees of freedom [8–14]. For instance, a singlet-triplet spin qubit encoded in the two-electron Hilbert space allows fast operation without the need of high-frequency microwave pulses. In addition, it has a good compatibility with fast and high-fidelity singlet-triplet based readout compared to spin-1/2 qubits [15,16]. Nevertheless, the qubit control fidelity in this system is yet below the fault-tolerant threshold of about 99% [17].

The singlet-triplet spin qubit makes use of the exchange interaction and therefore susceptible to charge noise, in addition to magnetic fluctuations due to nuclear spins in the host semiconductor material [8,9]. The magnetic noise can be most efficiently suppressed by the use of silicon-based material with reduced nuclear spin carrying isotopes [2–7,18,19]. The influence of charge noise, on the other hand, can be addressed by several approaches; symmetric operation [19,20], resonant operation in a large field gradient [21]. The resonant operation in a GaAs-based device has led to the highest control fidelity of 98.6 %, while it still suffers from the nuclear magnetic fluctuation and the detuning charge noise due to operation at a large detuning [21]. Here we show that by combining these approaches with silicon-QDs the exchange-based qubit control fidelity can reach a fault-tolerant level [17] as demonstrated through randomized benchmarking.

In this Letter, we operate and characterize a resonantly driven singlet-triplet spin qubit in silicon (Si). The spin qubit is defined by the two-electron anti-parallel spin states $|\widetilde{\downarrow\uparrow}\rangle$ and $|\widetilde{\uparrow\downarrow}\rangle$ in an exchange coupled DQD under a large magnetic field gradient. The tilde indicates the hybridization of the spin eigenstates without the exchange interaction $|\downarrow\uparrow\rangle$ and $|\uparrow\downarrow\rangle$ [5]. The coherent driving of the qubit can be performed by modulating the exchange interaction at the frequency of qubit energy splitting which is typically below 1 GHz. This is much lower in frequency than what is required to drive a spin-1/2 qubit (for example, ~14 GHz at a magnetic field of 0.5 T) and a standard arbitrary waveform generator (AWG) can be used for the resonant pulse generation. The relatively low-frequency control may facilitate the application of control pulses in a scalable manner. The qubit has a coherence time and a control fidelity comparable to those reported for spin-1/2 qubits in similar isotopically natural Si materials [5,6,18].

Figure 1(a) shows a scanning electron microscope image of our Si/SiGe QD device. Three layers of overlapping aluminium gates [22] deposited on top of an isotopically natural Si/SiGe heterostructure are used to form a DQD (Fig. 1(b)). The aluminium gates are insulated from each other by a layer of thin native aluminium oxide [23]. A cobalt micromagnet is placed on top of the QD array to induce a local magnetic field gradient. A nearby sensor QD coupled to a radio-frequency tank circuit allows rapid measurement of the charge configuration [24]. All measurements were performed in a dilution refrigerator with a base electron temperature $T_\text{e} \sim 40$ mK. An in-plane external magnetic field $B_\text{ext} = 0.5$ T is applied using a superconducting magnet.

The number of electrons inside the QD is controlled by the plunger gates P1 and P2, while the barrier gate B2 provides a control over the tunnel coupling $t_\text{C}$ between the right and left QDs. The qubit is operated in the (1,1) charge configuration where the numbers ($n_\text{L}$, $n_\text{R}$) represent the charge occupation of the left ($n_\text{L}$) and right ($n_\text{R}$) QDs. Gates P1, P2, and B2 are connected to an AWG (Tektronix AWG5208) running at a sampling rate of 1 GSa/s. The a.c. voltage pulses which modulate the exchange interaction are directly generated by the AWG. The electric-dipole spin resonance (EDSR) pulses used for spin initialization are generated by a Keysight E8267D microwave vector signal generator. The microwave signal is I/Q modulated by another Tektronix AWG5208 unit.

Our qubit is operated in the (1,1) charge configuration and the qubit state consists of two antiparallel eigenstates of the two-spin system, $|\widetilde{\uparrow\downarrow}\rangle$ and $|\widetilde{\downarrow\uparrow}\rangle$, under a finite exchange interaction $J$. The energy diagram of unpolarized spin states of a DQD is shown in Fig.

1(c). The inhomogeneous dephasing time $T_2^*$ would be largest at around $\varepsilon = 0$, where the detuning susceptibility of $J$, $|dJ/d\varepsilon|$ is minimized [19,20]. However, at the exact symmetric operation point, the qubit control speed would be lowest. Therefore, to increase the qubit control speed, we operate our qubit at the largest $\varepsilon$ where $T_2^*$ is not significantly degraded by charge noise unless noted. When driven, the rotating frame Hamiltonian at the drive frequency can be written as $H_{\text{RWA}} = hf_R(\cos\phi\,(\sigma_x/2) + \sin\phi\,(\sigma_y/2)) + (\sqrt{J_0^2 + \Delta E_z^2} - hf_{\text{a.c.}})(\sigma_z/2)$. Here, $h$ is the Planck's constant, $J_0$ is the mean value of exchange energy, $\phi$ is the phase of the a.c. drive, $f_{\text{a.c.}}$ is the frequency of resonant drive, $\Delta E_z$ is the Zeeman energy difference between the two QDs, and $f_R$ is half the a.c. modulation amplitude at $f_{\text{a.c.}}$ perpendicular to the quantization axis of the resonant qubit. As in the standard spin resonance experiments, two-axis universal control can be implemented by modulating $\phi$. Figure 1(d) shows a charge stability diagram measured as a function of the plunger gate voltages $V_{P1}$ and $V_{P2}$. The detuning is defined as $(\delta V_{P1}, \delta V_{P2}) = (1, -1.1)\delta\varepsilon$ and its origin is at around the center of (1,1) charge configuration.

We now proceed to demonstrate the basic operations of our resonantly driven singlet-triplet qubit. Figure 2(a) shows the measurement sequence. First, the electron spin in the right QD is initialized to spin-down state near the (1,0)-(0,1) transition [25]. We then initialize the left spin by spin-selective tunneling at the (0,1)-(1,1) boundary. Next, a gate voltage pulse is applied to push the electrons deep into the Coulomb blockade and an EDSR pulse is applied to rotate the $|\downarrow\downarrow\rangle$ state to $|\downarrow\uparrow\rangle$. $J$ is turned on by a 0.07 V square voltage pulse to the B2 gate. The gate voltage pulse has a 20 nsec rise time in order to adiabatically turn on $J$ with respect to $\Delta E_z$. After the initialization process, we perform the qubit operation by applying a.c. voltage pulses to the B2 gate. Finally, $J$ is turned off and we perform single-shot energy-selective readout of the left spin near the (0,1)-(1,1) state boundary. This maps out $|\downarrow\uparrow\rangle$ to spin-down and $|\uparrow\downarrow\rangle$ to spin-up readout outcomes [26]. We collect such 400 to 1,000 single-shot outcomes to obtain the probability of finding $|\uparrow\downarrow\rangle$. This readout protocol is robust against the large $\Delta E_z$, but the Pauli spin blockade will also work using the latched readout mechanism [15,16] or the shelving process [21,27].

Figure 2(b) shows measured exchange Rabi chevron pattern, which displays the qubit resonance frequency $\sqrt{\Delta E_z^2 + J_0^2}/h = 351$ MHz. No significant Rabi oscillation decay is observed for the a.c. pulse duration used here. We obtain an exchange Rabi frequency $f_R \sim 4$ MHz, which is comparable to the typical values for ESDR in similar devices [2,5,18].

Here the maximum a.c. voltage amplitude is limited by the experimental setup. Figure 2(c) shows Rabi oscillation measured for a longer burst time at the resonance condition. From this measurement, we obtain a 1/e Rabi oscillation decay time $T_R \sim 6\,\mu s$, which is long enough to allow for high-fidelity qubit control. Figure 2(d) shows the a.c. voltage amplitude dependence of the Rabi oscillations. Figure 2(e) shows the Rabi frequencies extracted from the data in Fig. 2(d). The Rabi frequency changes linearly in the measured range of the a.c. voltage pulse amplitude, indicating that the qubit is in the regime where $J$ changes linearly with $\delta V_{B2}$.

To access $T_2^*$ and the influence of charge noise, we perform Ramsey interferometry experiments for various detuning $\varepsilon$ (Fig. 3(a)). The Ramsey fringe measured at each $\varepsilon$ is fit by a Gaussian decay to extract the dephasing rate $(T_2^*)^{-1}$ (Figs. 3(b)-(e)). The dephasing rate turns out to vary only slightly within a relatively large window $-10\,\text{mV} \lesssim \varepsilon \lesssim 20\,\text{mV}$. The weak $\varepsilon$ dependence of $T_2^*$ around the symmetric operation point indicates that $T_2^*$ is not limited by the detuning noise. In addition, $T_2^*$ obtained around the symmetric operation point is consistent with $T_2^* \sim 1.8\,\mu s$ measured for the right and left spin-1/2 qubits in a more weakly coupled condition using EDSR (data not shown). We therefore conclude that our resonantly driven qubit is limited by the 4.7% $^{29}$Si nuclei in the isotopically natural Si quantum well rather than the charge noise. We note that the nuclei-induced $T_2^*$ obtained here are 3 to 4 times longer than the value previously reported for a singlet-triplet qubit in a similar material ($T_2^* \sim 0.36\,\mu s$ in Ref. [9]), perhaps due to the difference in the data acquisition time [28]. Far away from the symmetric operation point, we approach the inter-dot transition and the detuning noise starts to dominate the dephasing. For the Rabi oscillation and randomized benchmarking measurements, we choose the operation point at $\varepsilon = 20\,\text{mV}$ to increase $f_R$. This operation point barely affects $T_2^*$ while enabling roughly 2 times faster $f_R$ for the same a.c. voltage amplitude.

Finally, the qubit performance is characterized by randomized benchmarking [29]. Here, we twirl the qubit state in the subspace spanned by $|\widetilde{\downarrow\uparrow}\rangle$ and $|\widetilde{\uparrow\downarrow}\rangle$ and the performance of single-qubit control is evaluated. The 24 single-qubit Clifford gates are decomposed into rotations around x- and y-axes as in Ref. [30], which results in 1.875 single gates on average per one Clifford gate. We measure the sequence fidelities for both recovery Clifford gates to result in $|\widetilde{\downarrow\uparrow}\rangle$ and $|\widetilde{\uparrow\downarrow}\rangle$ to remove the offset error. Figure 4 shows the measured sequence fidelity decay as a function of the number of Clifford gates applied. From the exponential decay of the sequence fidelity, we extract a depolarizing parameter $p = 0.985 \pm 0.0009$, which results in a Clifford gate fidelity $F_\text{C} = 99.2 \pm 0.045$ % and single gate fidelity $F_\text{single} = 99.6 \pm 0.024$ %. The obtained fidelity is the highest reported for singlet-triplet spin qubit and satisfies the threshold for surface code quantum error correction [17].

In conclusion, we have demonstrated operation and fidelity benchmark of a resonantly driven singlet-triplet qubit in Si. It provides an alternative operation mode of high-fidelity spin qubits in Si. We anticipate that the performance of the qubit will be improved by using isotopically enriched $^{28}$Si because $T_2^*$ is currently limited by the nuclear magnetic noise. The same resonant control technique can be applied to an array of spin-1/2 qubits to implement a SWAP gate (with additional phase calibrations), initialization and measurement of spins not directly connected to the reservoirs. Indeed, during the preparation of the manuscript, we became aware of the application of a similar technique to transfer information of spin-1/2 qubits [31].

Acknowledgements

We thank the Microwave Research Group in Caltech for technical support. This work was supported financially by Core Research for Evolutional Science and Technology (CREST), Japan Science and Technology Agency (JST) (JPMJCR15N2 and JPMJCR1675) and MEXT Quantum Leap Flagship Program (MEXT Q-LEAP) grant No. JPMXS0118069228. K.T. acknowledges support from JSPS KAKENHI grant No. JP17K14078. A.N. acknowledges support from JSPS KAKENHI grant No. 19K14640. T.N. acknowledges support from RIKEN Incentive Research Projects. S.T. acknowledges support from JSPS KAKENHI grant Nos. JP26220710 and JP16H02204.

# Figures and tables

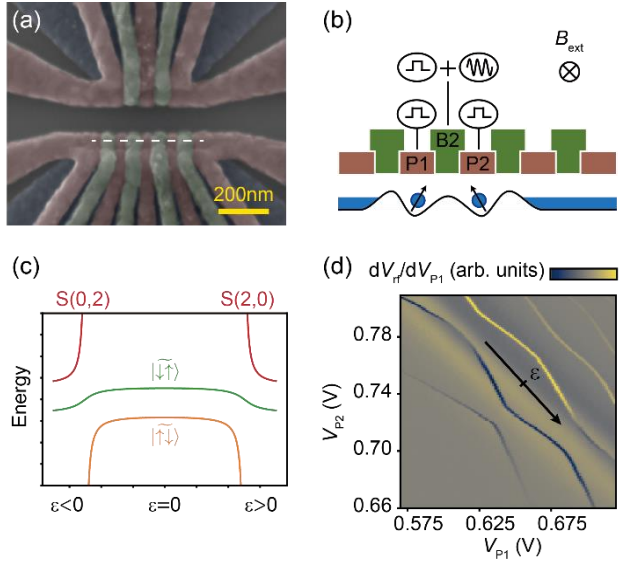

**Figure 1.** (a) False colored scanning electron microscope image of the device. Three layers of overlapping aluminium gates are used to control the confinement potential. The screening gates (blue) are used to restrict the electric field of the plunger (red) and barrier (green) gates. (b) Schematic of device geometry and measurement setup. The device geometry shows a line cut along the white dashed line in Fig. 1(a). Three gates labelled as P1, P2, and B2 are mainly used to control the DQD confinement. (c) Energy diagram of two-electron unpolarized spin states. (d) Charge stability diagram measured as a function of gate voltages $V_{P1}$ and $V_{P2}$. The variation of background signal is caused by the Coulomb oscillation of the radio-frequency sensor QD. The tick of the detuning axis indicates $\varepsilon = 0$.

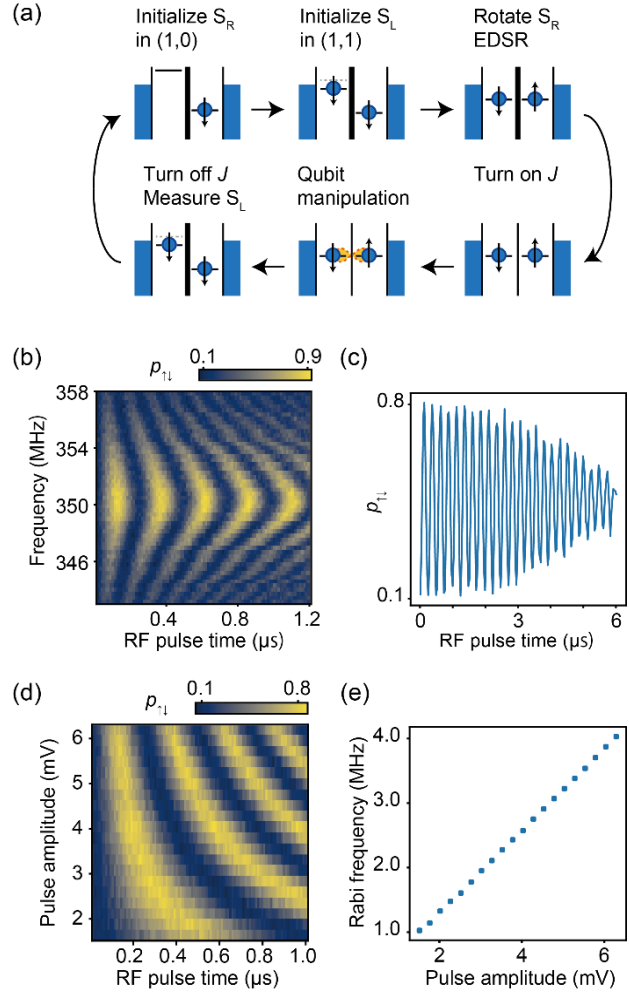

**Figure 2.** (a) Measurement sequence of the resonantly driven spin qubit. S$_L$ and S$_R$ refers to the left and right spin, respectively. (b) Rabi chevron pattern measured at the a.c. pulse amplitude of 6.3 mV. (c) The Rabi oscillation measured for a longer RF pulse duration. The Rabi frequency is set at the center resonance frequency $f = 351$ MHz. (d) Rabi oscillation power dependence. (e) Rabi frequencies extracted from the power dependence measurement. Each of the Rabi oscillations in Fig. 2(d) is fit by a sine curve $p_{\uparrow\downarrow}(t) = A\sin(2\pi f_R t - \pi/2) + B$, where $A$ and $B$ are the constants to account for the readout fidelities and $f_R$ is the Rabi frequency.

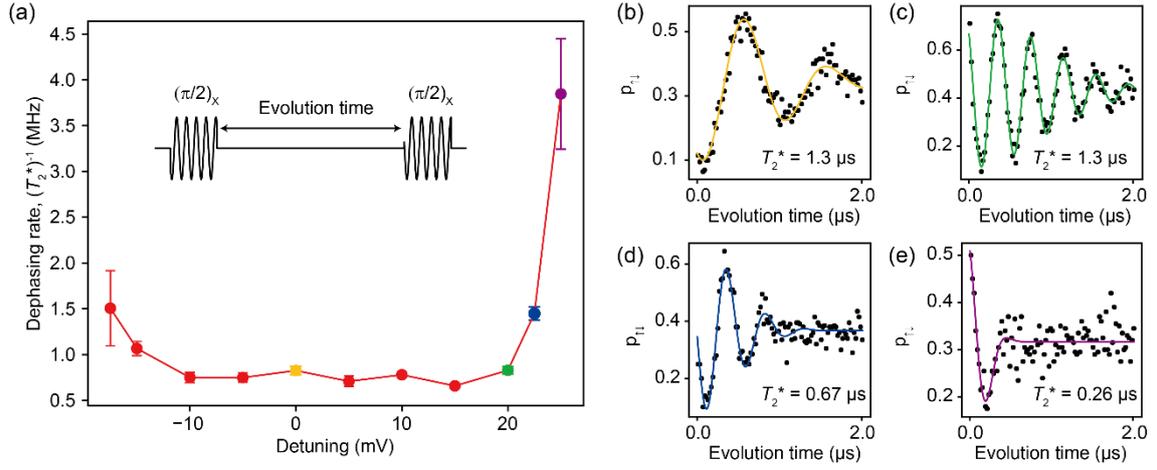

Figure 3. Detuning dependence of the phase coherence time. (a) Detuning dependence of the phase coherence time measured by Ramsey interferometry. The error bars represent one sigma from the mean. The inset schematic shows the measurement sequence of the Ramsey interferometry. First we apply π/2 pulse and wait for some time. Finally, the phase accumulated during the waiting time is projected to z-axis by another π/2 pulse. (b)-(e) Ramsey fringes measured at various detuning conditions. Each curve is fit by a Gaussian decaying oscillation and $T_2^*$ is extracted. The detuning values are 0 mV for (b), 20 mV for (c), 22.5 mV for (d), and 25 mV for (e).

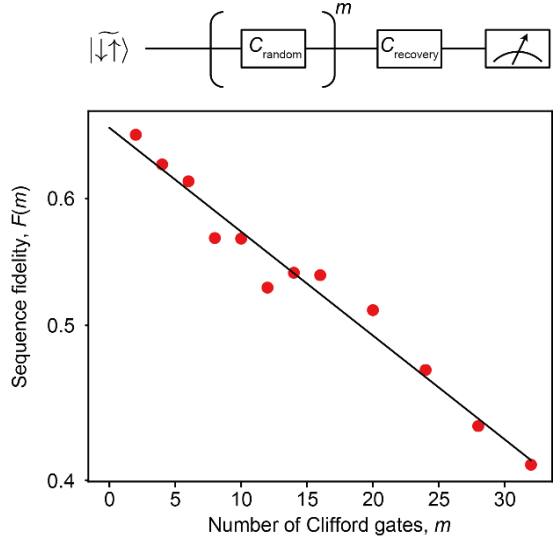

**Figure 4.** Randomized benchmarking measurement. The sequence fidelity $F(m)$ is defined as $F(m) = p_{\uparrow\downarrow}^{|\widetilde{\uparrow\downarrow}\rangle}(m) - p_{\uparrow\downarrow}^{|\widetilde{\downarrow\uparrow}\rangle}(m)$, where $p_{\uparrow\downarrow}^{|\widetilde{\uparrow\downarrow}\rangle}(m)$ ($p_{\uparrow\downarrow}^{|\widetilde{\downarrow\uparrow}\rangle}(m)$) is the probability of finding an $|\widetilde{\uparrow\downarrow}\rangle$ state after applying the recovery Clifford gate designed to result in an ideal outcome $|\widetilde{\uparrow\downarrow}\rangle$ ($|\widetilde{\downarrow\uparrow}\rangle$). The decay curve is fit by an exponential decay $F(m) = Vp^m$, where $V$ is the visibility. We obtain $V = 0.665 \pm 0.009$ and $p = 0.985 \pm 0.0009$ from the fit. The fitting errors represent one sigma from the mean.